\documentclass[acmsmall,screen]{acmart} 

\AtBeginDocument{%
  \providecommand\BibTeX{{%
    \normalfont B\kern-0.5em{\scshape i\kern-0.25em b}\kern-0.8em\TeX}}}

\setcopyright{acmcopyright}
\copyrightyear{2022}
\acmYear{2022}
\acmDOI{XXXXXXX.XXXXXXX}

\acmJournal{JACM}
\acmVolume{37}
\acmNumber{4}
\acmArticle{111}
\acmMonth{8}

\usepackage{bm}
\usepackage{array}
\usepackage{color}
\usepackage{algorithm}
\usepackage{algorithmic}
\usepackage{multirow}
\usepackage{makecell}
\usepackage{enumerate}
\usepackage{enumitem}
\allowdisplaybreaks[3]

\begin{document}

\title{Dual Preference Distribution Learning for Item Recommendation}

\author{Xue Dong}
\email{dongxue.sdu@gmail.com}
\affiliation{%
  \institution{School of Software, Shandong University}
  \city{Jinan}
  \country{China}
  \postcode{250101}
}

\author{Xuemeng Song}
\authornote{Corresponding author.}
\email{sxmustc@gmail.com}
\affiliation{%
  \institution{School of Computer Science and Technology, Shandong University}
  \city{Qingdao}
  \country{China}
  \postcode{266237}
}

\author{Na Zheng}
\email{zhengnagrape@gmail.com}
\affiliation{%
  \institution{Institution of Data Science, National University of Singapore}
  \country{Singapore}
}

\author{Yinwei Wei}
\email{weiyinwei@hotmail.com}
\affiliation{%
  \institution{School of Computing, National University of Singapore}
  \country{Singapore}
}

\author{Zhongzhou Zhao}
\email{zhongzhou.zhaozz@alibaba-inc.com} 
\affiliation{%
  \institution{DAMO Academy, Alibaba Group}
  \city{Zhejiang}
  \country{China}
  \postcode{311121}
}

\begin{abstract}
Recommender systems can automatically recommend users with items that they probably like. The goal of them is to model the user-item interaction by effectively representing the users and items. Existing methods have primarily learned the user's preferences and item's features with vectorized embeddings, and modeled the user's general preferences to items by the interaction of them. In fact, users have their specific preferences to item attributes and different preferences are usually related. Therefore, exploring the fine-grained preferences as well as modeling the relationships among user's different preferences could improve the recommendation performance.
Toward this end, we propose a dual preference distribution learning framework (DUPLE), which aims to jointly learn a general preference distribution and a specific preference distribution for a given user, where the former corresponds to the user's general preference to items and the latter refers to the user's specific preference to item attributes.
Notably, the mean vector of each Gaussian distribution can capture the user's preferences, and the covariance matrix can learn their relationship. Moreover, we can summarize a preferred attribute profile for each user, depicting his/her preferred item attributes. We then can provide the explanation for each recommended item by checking the overlap between its attributes and the user's preferred attribute profile.
Extensive quantitative and qualitative experiments on six public datasets demonstrate the effectiveness and explainability of the DUPLE method.
\end{abstract}


\setcopyright{acmcopyright}
\acmJournal{TOIS}
\acmYear{2022} \acmVolume{1} \acmNumber{1} \acmArticle{1} \acmMonth{1} \acmPrice{15.00}\acmDOI{10.1145/3565798}


\maketitle

\section{Introduction}
Recommender systems that aim to recommend users items that they probably like have been attracting increasing research attention~\cite{Rendle10,McAuleyTSH15,Wang00LC19,WangJZ0XC20}.
Mainstream approaches target at learning user and item embeddings to represent the user preferences and item properties, respectively. They then predict the user's preference to an item according to certain interaction score between their embeddings~\cite{KorenBV09,HeLZNHC17,TayTH18,Wang0WFC19}. This kind of preference reflects the user's overall judgment to the item, termed as the user's general preference.
Recently, several research attempts~\cite{ZhangL0ZLM14,PanLLZ20} further explored the user's fine-grained attitudes to the item's side information (e.g., attributes), termed as the user's specific preference, to improve the recommendation performance and explainability. They typically adopt the attention mechanism to fuse the attribute embeddings as the extension of the item embedding, whereby the attention weights can be as the proxies of the user's attitudes to different item attributes. Despite of their compelling success, existing approaches overlook the relationships among the user's multiple preferences, which are also necessary to understand the user preferences and improve the recommendation performance. For example, if a user has watched many action movies, most of which are played by the actor Jackie Chan, then we learn that the user's preferences to \textit{action} and \textit{Jackie Chan} are positively related and we can recommend the user with other movies played by Jackie Chan. 


Towards this end, we propose a dual preference distribution learning framework for the explainable item recommendation, termed DUPLE, which aims to jointly learn a general preference distribution and a specific preference distribution for a given user, where the former corresponds to the user’s general preference to items, while the latter refers to the user’s specific preference to item attributes. In particular, we adopt the widely-used Gaussian distribution as the format of these two preference distributions. In this manner, the mean vector that locates the distribution can capture the user's main preferences, while the covariance matrix reflected the relationships between latent variables can model the relationships among the user's different preferences. Moreover, tracing to the user's specific preference distribution, we can summarize the user's preferences to item attributes and perform the explainable recommendation.

Specifically, as illustrated in Fig.~\ref{figure_model}, DUPLE consists of three key components: the \textit{general preference learning}, \textit{specific preference learning}, and \textit{explanation production}. In the first component, we introduce a parameter construction module to learn the essential parameters (i.e., the mean vector and covariance matrix) of the user's general preference distribution based on the user embedding. In the second component, we design a general-specific transformation module to infer the user's specific preference distribution from the general one. The underlying philosophy is that we bridge the gap between the user's general preference to items and specific preference to item attributes with the help of the projection between the item embedding and attribute embeddings, transforming the parameters of the general preference distribution into their specific counterparts. We predict the user rating to an item by the probability densities of the item in both the user’s general and specific preference distributions. Ultimately, once the specific preference distribution has been learned, in the third component, we can summarize a preferred attribute profile to store the user's preferences to item attributes, and provide explanation for each recommended item by checking the overlap between the item's attributes and user's preferred attribute profile. We summarize our main contributions as follows:

\begin{figure*}[!t]
	\centering
	\includegraphics[width=\textwidth]{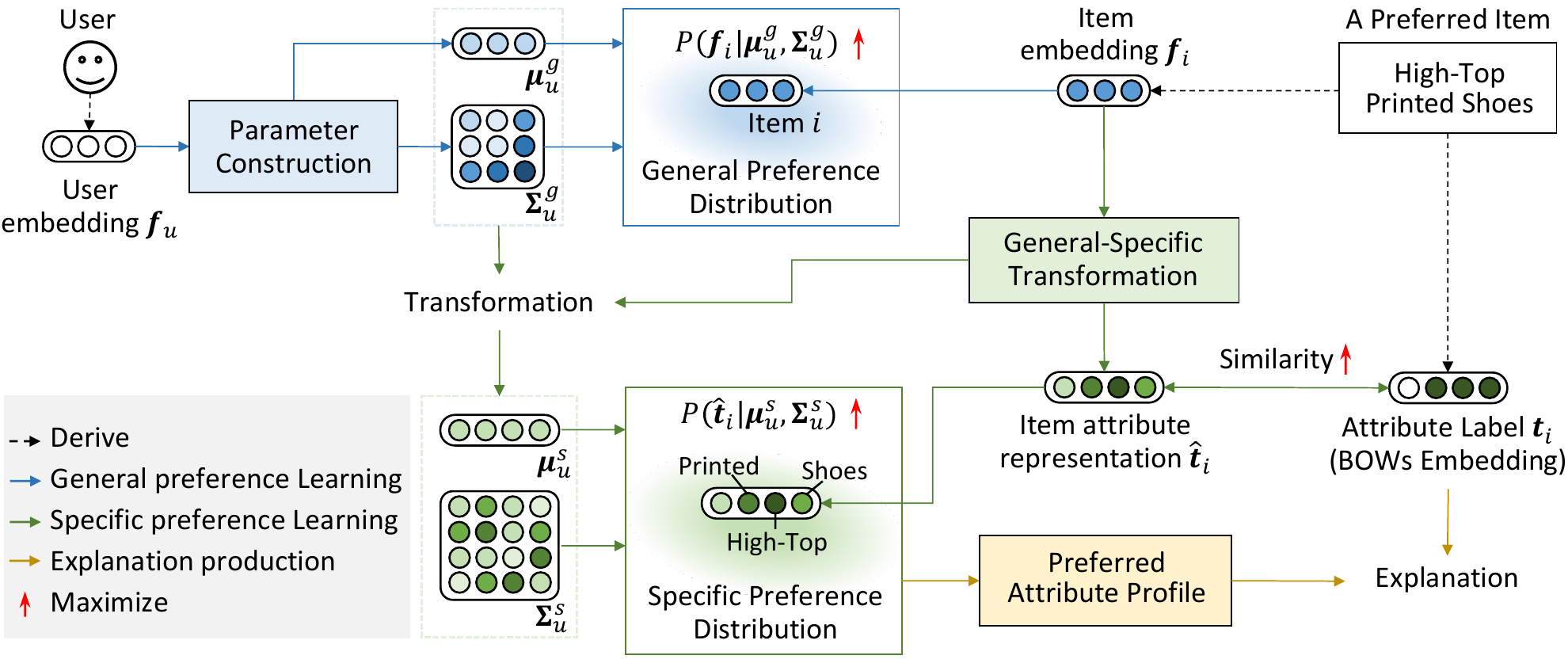}
	\caption{Illustration of the proposed dual preference distribution learning framework (DUPLE) for the explainable item recommendation, which jointly learns the user's preferences to general item (blue flow) and specific attribute (green flow) for the better recommendation. } 
	\label{figure_model}
\end{figure*}

\begin{itemize}
    \item We propose a dual preference distribution learning framework (DUPLE) that captures both the user's overall judgment to items (i.e., the general preference) and fine-grained attitude to item attributes (i.e., the specific preference) with Gaussian distributions. Benefited from the covariance matrix of the Gaussian distribution, we can capture the relationships between the user's different preferences for the better recommendation. Moreover, by checking the user's specific preference to item attributes, we can summarize a preferred attribute profile for the user and explain why recommend the item to the user.
    \item We design a general-specific transformation to derive the user's specific preference to item attributes from the general preference. This allows the knowledge learned by both the general and specific preferences to be mutually referred.
    \item We conduct extensive experiments on six version datasets derived from the Amazon Product~\cite{McAuleyTSH15} and MovieLens~\cite{HarperK16}, and the results demonstrate the superiority and explainability of our proposed DUPLE method over several state-of-the-art methods. We will release our codes to facilitate other researchers.
\end{itemize}

\section{Related work}
In this section, we briefly introduce traditional recommender systems in Subsection~\ref{section_trad_rs} and explainable recommender systems in Subsection~~\ref{section_explain_rs}.
\subsection{Recommender Systems}
\label{section_trad_rs}
Initial researches utilize the Collaborative Filtering techniques~\cite{SuK09} to capture the user's preferences from the interacted relationships between users and items. Matrix Factorization (MF) is the most popular method~\cite{LeeS00, SalakhutdinovM07, KorenBV09}. It focuses on factorizing the user rating matrix into the user matrix and item matrix and predicting the 
user-item interaction by the similarity of their representations. Considering that different users may have different rating habits, Koren~et~al.~\cite{KorenBV09} introduced the user and item biases into the matrix factorization, achieving a better performance. Several other researchers argued that different users may have similar preferences to items. Consequently, Yang~et~al.~\cite{YangWCSW20} and Chen~et~al.~\cite{ChenWZYMR20} clustered the users into several groups according to their historical item interactions and separately captured the common preferences of users in each group. Several approaches introduce to utilize probabilistic distribution to model the uncertainty of the user preferences~\cite{LiS17, LiangKHJ18, ShenYLZZLX21}.
Recently, due to the great performance of graph convolutional networks, many approaches have resorted to constructing a graph of users and items according to their historical interactions, and exploring the high-order connectivity from user-item interaction~\cite{Wang0WFC19, TaoWWHHC20,Wang00LC19,WangX000C20,WangJZ0XC20, 0008T0ZNY21}. For example, Wang~et~al.~\cite{Wang0WFC19} served users and items as nodes and their interaction histories as edges between nodes. And then, they proposed a three-layer embedding propagation to propagate messages from items to user and then back to items. Wang~et~al.~\cite{WangJZ0XC20} designed the intent-aware interaction graph that disentangles the item representation into several factors to capture user's different intents.

Beyond directly learning the user's preferences from the historically user-item interactions, other researchers began to leverage the item's rich context information to capture the user's detailed preferences to items. Several of them incorporated the visual information of items to improve the recommendation performance~\cite{McAuleyTSH15, HeM16, YangSFWDN21}. For example, He~et~al.~\cite{HeM16} enriched the item's representation by extracting the item's visual feature by a CNN-based network from its image and added it into the matrix factorization. Yang~et~al.~\cite{YangSFWDN21} further highlighted the region of the item image that the user is probably interested in. Some other researches explore the item attributes~\cite{ZhangL0ZLM14, BalogRK21, PanLLZ20} or user reviews~\cite{ChenW0WBWC19, XianFZGCHG0MMZ20, TalLHYA21} to learn the user's preferences to the item's specific aspects. For example, Pan~et~al.~\cite{PanLLZ20} utilized the attribute representations as the regularization of learning the item's representation during modeling the user-item interaction. In this manner, the user's preferences to attributes can be tracked by the path of the user to item and then to attributes. And Chen~et~al.~\cite{ChenW0WBWC19} proposed a co-attentive multi-task learning model for recommending user items and generating the user's reviews to items. Besides, the multi-modal data has been proven to be important for the recommendation~\cite{MMGCN2019MM, DongWSDN20, LeiLZ0TLM21, QiuWCYH21}. For example, Wei~et~al.~\cite{MMGCN2019MM} constructed a user-item graph on each modality to learn the user’s modal-specific preferences and incorporated all the preferences to predict the user-item interaction.

Although these above approaches are able to capture the user's preferences to items or specific contents of the item (e.g., attributes), they fail to model the relationships among the user's different preferences, which is beneficial for the better recommendation. In this work, we proposed to capture the user's preferences with the multi-variant Gaussian distribution, and model the user-item interaction with the probability density of the item in the user's preference distribution. In this manner, the relationships among the user's preferences can be captured by the covariance matrix of the distribution.

\subsection{Explainable Recommender Systems}
\label{section_explain_rs}
The recommender systems trained by deep neural networks are perceived as a black box only able to predict a recommendation. Thus, to make the recommendation more transparent and trustworthy, explainable recommender systems~\cite{ZhangL0ZLM14,ChenZHAXH18} are therefore gaining popularity, which focus on what and why to recommend an item. Initial approaches can provide rough reasons of recommending an item based on the similar items or users~\cite{SchaferKR99, SharmaC13}. Thereafter, researchers attempted to seek more explicit reasons provided to users, e.g., explain recommendations with item attributes, reviews, and reasoning paths. 

\textbf{Explanation with Item Attributes.} 
This group of explainable recommender systems consider that the user likes an item may be caused by its certain attributes, e.g., ``you may like Harry Potter because it is an adventure movie''. Thus, mainstream approaches in this research line have been dedicated to bridging the gap between users and attributes~\cite{ChenZHAXH18, HouYWY19, Yang0WMF0C19, PanLLZ20}. In particular, Wang~et~at.~\cite{Wang0FNC18} employed a tree-based model to learn explicit decision rules from item attributes, and designed an embedding model to generalize to unseen decision rules on users and items. Benefited from the attention mechanism, several researchers~\cite{ChenZHAXH18, PanLLZ20} learned the item embedding with the fusion of its attribute embeddings. By checking the attention weights, these methods can infer how each attribute causes the high/low rating score. 

\textbf{Explanation with Reviews.}
These methods leverage the review as the extra information to the user-item interaction, which can infer the user's attitude towards one item~\cite{SeoHYL17, CostaODL18, LuDS18WWW, WuQLWZL19, ChengCZKK19}. In particular, several approaches aggregate review texts of users/items and adopt the attention mechanism to learn the users/items embeddings~\cite{SeoHYL17, WuQLWZL19, LuDS18WWW, GuanCHZZPC19}. Based on the attention weights, the model can highlight the words in reviews as explanations. 
Different from highlighting the review words as explanations, other researches attempt to automatically generate reviews for a user-item pair~\cite{LiWRBL17, LuDS18, CostaODL18, ChenW0WBWC19}. Specifically, Costa~et~al.\cite{CostaODL18} designed a character-level recurrent neural network, which generates review explanations for the user-item pair using long-short term memories.
Li~et~al.\cite{LiWRBL17} proposed a more comprehensive model to generate tips in review systems. Inspired by human information processing model in cognitive psychology, Chen~et~al.~\cite{ChenW0WBWC19} developed an encoder-selector-decoder architecture, which exploits the correlations between recommendation and explanation through co-attentive multi-task learning.

\textbf{Explanation with Reasoning Paths.}  
This kind of approaches construct a user-item interaction graph and aim to find a explicitly path on the graph that traces the decision-masking process~\cite{AiACZ18, Wang00LC19, HuangFQSLX19, XianFMMZ19, HeLZLW20, XianFZGCHG0MMZ20}. In particular, Ai~et~al.~\cite{AiACZ18} constructed a user-item knowledge graph of users, items, and multi-type relations (e.g., purchase and belong), and generated explanations by finding the shortest path from the user to the item. Wang~et~al.~\cite{Wang00LC19} proposed a Knowledge Graph Attention Network (KGAT) that explicitly models the high-order relations in the knowledge graph in an end-to-end manner. Xian~et~al.~\cite{XianFMMZ19} proposed a reinforcement reasoning approach over knowledge graphs for interpretable recommendation, where agent starts from a user and is trained to reach the correct
items with high rewards.
Further, considering that users and items have different intrinsic characteristics, He~et~al.~\cite{HeLZLW20} designed a two-stage representation learning algorithm for learning better representations of heterogeneous nodes. Yang~et~al.~\cite{YangD20} proposed a Hierarchical Attention Graph Convolutional Network (HAGERec) that involves the hierarchical attention mechanism to exploit and adjust the contributions of each neighbor to one node in the knowledge graph.

Despite of their achievements in the explainable recommendation, existing approaches are mainly discriminative methods to provide an explanation, i.e., they provide an explanation for a given user-item pair. However, in fact, users have their inherent preferences guiding them to select items. Therefore, we propose to mimic this practical manner that first summarize the user's preferences and then explain the recommendation by the overlap between the user's preferences and item properties.

\section{The Proposed DUPLE Model}
To improve the readability, we declare the notations used in this paper. We use the squiggled letters (e.g., $\mathcal{X}$) to represent sets. The bold capital letters (e.g., $\bm{X}$) and bold lowercase letters (e.g., $\bm{x}$) represent matrices and vectors, respectively. Let the non-bold letters (e.g., $x$) denote scalars. The notations used in this paper are summarized in Table~\ref{table_notations}.

We now present our proposed dual preference distribution learning framework (DUPLE) for the explainable item recommendation, which is illustrated in Fig.~\ref{figure_model}. 
It is composed of three key components: 1) general preference learning, where a parameter construction module is proposed to construct the essential parameters of the user's general preference distribution; 2) specific preference learning, where we propose a general-specific transformation module to learn the user's specific preference distribution by transforming the parameters of the general one into their specific counterparts; and 3) the explanation production that summarizes the preferred attribute profile for the user and explains why recommending an item for a user from the item attribute perspective.
In the rest of this section, we first briefly define our explainable recommendation problem in Subsection~\ref{section_problem_definition}. Then, we detail the three key components in Subsections~\ref{section_general_preference}, \ref{section_specific_preference}, and \ref{section_explanation}, respectively. Finally, we illustrate the model optimization in Subsection~\ref{section_interaction}.

\subsection{Problem Definition}
\label{section_problem_definition}
Without losing generality, suppose that we have a set of users~$\mathcal{U}$, a set of items~$\mathcal{I}$, and a set of item attributes~$\mathcal{A}$ that can be applied to describe all the items in $\mathcal{I}$. Each user~$u \in \mathcal{U}$ is associated with a set of items~$\mathcal{I}_u$ that the user historically likes. 
Each item~$i \in \mathcal{I}$ is annotated by a set of attributes~$\mathcal{A}_i$. Following mainstream recommender models~\cite{LeeS00, HeLZNHC17, WangJZ0XC20}, we describe
a user~$u$ (an item~$i$) with an embedding vector $\bm{f}_u \in \mathbb{R}^{D}$ ($\bm{f}_i \in \mathbb{R}^{D}$), where $D$ denotes the embedding dimension. Besides, to describe the item with its attributes, we represent the item~$i$ by an attribute embedding, i.e., the bag-of-words embedding of its attributes~$\bm{t}_i \in \mathbb{R}^{|\mathcal{A}|}$, where the $j$-th elements $\bm{t}_i^j=1$ refers to that the item has the $j$-th attribute in~$\mathcal{A}$. 

\textbf{Inputs:} The inputs of the dual preference distribution learning framework (DUPLE) consist of 3 parts: the set of users~$\mathcal{U}$, the set of items~$\mathcal{I}$, and the set of attributes~$\mathcal{A}_i$ of the item~$i \in \mathcal{I}$.

\textbf{Outputs:} 
Given a user~$u$ and item~$i$ with its attributes~$\mathcal{A}_i$, DUPLE predicts the preference~$p_{ui}$ from both the general item and specific attribute perspectives as follows, 
\begin{equation}
	\label{equation_pui}
	p_{ui} = \lambda p_{ui}^g+(1-\lambda) p_{ui}^s, \\
\end{equation}
where $p_{ui}^g$ and $p_{ui}^s$ are the general and specific preferences of the user~$u$ to the item~$i$ (will be introduced in Subsection~\ref{section_general_preference} and~\ref{section_specific_preference}), respectively. 
$\lambda \in [0,1]$ is a hyper-parameter for adjusting the trade-off between the two terms. 
Besides, DUPLE can summarize a preferred attribute profile $\mathcal{A}_u$ for the user~$u$ and provide the explanation for recommending an item~$i$ with the form of ``you may like $\mathcal{A}_u \cap \mathcal{A}_i$ of the item''.

\begin{table}[t]\small
	\centering
	\renewcommand{\arraystretch}{1.2}
	\caption{Summary of the Main Notations.}
	\label{table_notations}
	\setlength{\tabcolsep}{2mm}{
		\begin{tabular}{c|l}	
			\hline
			Notation & Explanation \\
			\hline\hline
			$\mathcal{U}$, $\mathcal{I}$, $\mathcal{A}$ & The sets of users, items, and attributes, respectively. \\
			
			$\mathcal{I}_u$ & The set of historical interacted items of the user~$u \in \mathcal{U}$.\\
			
			$\mathcal{A}_i$ & The set of attributes of the item~$i \in \mathcal{I}$.\\
			$\mathcal{A}_u$ & The summarized preferred attribute profile of the user~$u$.\\
			$\mathcal{D}$ & The training set of triplet $(u, i, k), u\in\mathcal{U}, i\in\mathcal{I}_u, k\notin \mathcal{I}_u$.\\
			$\bm{f}_u$, $\bm{f}_i$ & Embeddings of the user~$u$ and item~$i$, respectively. \\
			
			$\bm{t}_i$ & Bag-of-words attribute embedding of the item~$i$. \\
			
			$\bm{\mu}_u^g$, $\bm{\Sigma}_u^g$ & Mean vector and covariance matrix of the user $u$'s general preference distribution.\\
		
			$\bm{\mu}_u^s$, $\bm{\Sigma}_u^s$ & Mean vector and covariance matrix of the user $u$'s specific preference distribution.\\
			$p^g_{ui}$, $p^s_{ui}$ & The general and specific preferences of the user~$u$ to the item~$i$, respectively.\\
			$p_{ui}$ & The final preference of the user~$u$ to the item~$i$.\\
			
			$\bm{\Theta}$ & To-be-learned set of parameters.\\
			\hline
	\end{tabular}}
\end{table}

\subsection{General Preference Learning}
\label{section_general_preference}
We utilize a general preference distribution~$\mathcal{G}(\bm{\mu}_u^g, \bm{\Sigma}_u^g)$ for the user $u$ to capture his/her general preferences to items. More specifically, the mean vector~$\bm{\mu}_u^g$ refers to the center of the user $u$'s general preference, while the covariance matrix~$\bm{\Sigma}_u^g$ stands for the relationships among the latent variables affecting the user's preferences to items. 

Thus, the key of learning the user's general preference distribution is to construct the mean vector and covariance matrix. We design a parameter construction module to separately construct the mean vector and covariance matrix, as they are different in form and mathematical properties. In particular, given the user embedding~$\bm{f}_u$, we adopt one fully connected layer to derive the mean vector~$\bm{\mu}_u^g\in \mathbb{R}^{D}$ of the general preference distribution as follows,
\begin{equation}
\label{equation_mu}
\bm{\mu}_u^g = \bm{W}_{\mu}\bm{f}_u+\bm{b}_{\mu}, \\
\end{equation}
where $\bm{W}_{\mu}\in \mathbb{R}^{D\times D}$ and $\bm{b}_{\mu}\in \mathbb{R}^{D}$ are the non-zero parameters to map the user embedding.

The covariance matrix should be symmetric and positive semi-definite according to its mathematical properties. Therefore, it is difficult to construct it directly based on the user embedding. Instead, we propose to first derive a low-rank matrix $\bm{V}_u \in \mathbb{R}^{D\times D'}, D' < D$, from the user embedding as a bridge, and then construct the covariance matrix through the following equation,
\begin{equation}
\label{equation_sigma}
\bm{\Sigma}_u^g=\bm{V}_u\bm{V}_u^\mathsf{T},
\end{equation}
where $\bm{V}_u = [\bm{v}_u^1;\bm{v}_u^2;\cdots;\bm{v}_u^{D'}]$ is arranged by $D'$ column vectors, which can be derived by the column-specific transformation as follows,
\begin{equation}
\label{equation_equation_vuj}
\bm{v}_u^j = \bm{W}^j_{\Sigma}\bm{f}_u+\bm{b}^j_{\Sigma}, j=1,2,\cdots,D',
\end{equation}
where $\bm{W}^j_{\Sigma} \in \mathbb{R}^{D \times D}$ and $\bm{b}^j_{\Sigma}\in \mathbb{R}^{D}$ are the non-zero parameters to derive the $j$-th column of $\bm{V}_u$. These parameters and the user embedding $\bm{f}_u$ (i.e., non-zero vector) guarantee that the $\bm{v}_u^j$ is a non-zero vector. With the simple algebra derivation, the covariance matrix~$\bm{\Sigma}^g_u$ derived according to Eqn.~(\ref{equation_sigma}) is symmetric as ${\bm{\Sigma}_u^g}^\mathsf{T}=(\bm{V}_u\bm{V}_u^\mathsf{T})^\mathsf{T} = \bm{V}_u \bm{V}_u^\mathsf{T}=\bm{\Sigma}_u^g$, and positive semi-definite as $\bm{x}\bm{\Sigma}_u^g\bm{x}^\mathsf{T}=\bm{x}\bm{V}_u\bm{V}_u^\mathsf{T}\bm{x}^\mathsf{T}=||\bm{x}\bm{V}_u||^2\geq 0, \forall \bm{x} \neq \bm{0}, \bm{x} \in \mathbb{R}^{D}$. 

Based on the general preference distribution, we can derive the general preference of the user to one item. In particular, we adopt the probability density of the embedding~$\bm{f}_i$ of the item~$i$ (indicating the probability of the item belonging to the distribution) as the proxy of the general preference~$p^g_{ui}$ of the user~$u$ toward the item~$i$. Formally, we define 
$p^g_{ui}$ as follows, 
\begin{equation}
	\label{equation_p^g_ui}
		p^g_{ui} = P(\bm{f}_i|\bm{\mu}_u^g, \bm{\Sigma}_u^g) 
		= \frac{1}{\sqrt{2\pi|\bm{\Sigma}_u^g|}} \exp \big(-\frac{1}{2}(\bm{f}_i-\bm{\mu}_u^g)^\mathsf{T}({\bm{\Sigma}_u^g})^{-1}(\bm{f}_i-\bm{\mu}_u^g)\big),
\end{equation}
where $|\bm{\Sigma}^g_u|$ and $({\bm{\Sigma}_u^g})^{-1}$ are the determinant and inverse matrix of the covariance matrix $\bm{\Sigma}^g_u$ of the user's general preference distribution defined in Eqn.~(\ref{equation_sigma}), respectively. 
 
\subsection{Specific Preference Learning}
\label{section_specific_preference}
We learn the user's specific preference to item attributes from his/her general preference to items, with the help of bridging the gap between the items and their attributes. Specifically, we design a general-specific transformation module that predicts the item's attribute embedding from the item embedding. Thus, we can learn the user's specific preference distribution from the general preference distribution by transforming its mean vector and covariance matrix into their specific counterparts. Instead of introducing another branch to construct the specific preference distribution anew, i.e., learning its parameters $\bm{\mu}^s_u$ and $\bm{\Sigma}^s_u$ through the user embedding as similar to the parameter construction module, this design has the following two benefits.
(1)~The user's general preference toward an item often comes from his/her specific judgments toward the item's attributes. In light of this, it is promising to derive the specific preference distribution by referring to the general preference distribution. And (2)~learning the general and specific preferences of the user with one branch allows the knowledge learned by each component to be mutually referred. 

In particular, following the approach~\cite{PanLLZ20}, we adopt the linear mapping as the format of the general-specific transformation to predict the item~$i$'s attribute embedding~$\hat{\bm{t}}_i$ from the item embedding~$\bm{f}_i$ as follows,
\begin{equation}
\label{equation_transformation}
\hat{\bm{t}}_i=\bm{W}_t\bm{f}_i,
\end{equation}
where $\bm{W}_t \in \mathbb{R}^{|\mathcal{A}|\times D}$ is the parameter of the general-specific transformation. $|\mathcal{A}|$ is the total number of the item attributes.

We adopt the ground-truth attribute label~$\bm{t}_i$ of the item~$i$ to supervise the general-specific transformation learning. Specifically, inspired by the studies~\cite{SongFLLNM17, QiuHY21} that utilize the Bayesian Personalized Ranking (BPR) mechanism~\cite{RendleFGS09} to make the anchor more similar to its positive sample than a negative one, we adopt the following loss function to enforce the predicted attribute embedding~$\hat{\bm{t}}_i$ of the item~$i$ to be as close as to its ground truth attribute label vector~$\bm{t}_i$,
\begin{equation}
\label{equation_Lt}
\mathcal{L}_{t} = \sum_{i,j \in \mathcal{I}, i\neq j} -\log(\sigma(sim(\hat{\bm{t}}_i, \bm{t}_i)-sim(\hat{\bm{t}}_i, \bm{t}_j))),
\end{equation}
where $\bm{t}_j$ is the ground-truth attribute label vector of another item~$j$. $sim(\cdot)$ is the similarity between the two vectors.
Following the studies~\cite{HanWJD17, Wang00LC19}, we estimate the similarity between item~$i$'s attribute embedding and the ground-truth attribute label vector with their cosine similarity as: $sim(\bm{t}_i,\hat{\bm{t}}_i) = cos(\bm{t}_i,\hat{\bm{t}}_i)=\frac{\bm{t}_i^\mathsf{T} \hat{\bm{t}}_i}{||\bm{t}_i||||\hat{\bm{t}}_i||}$. By minimizing this objective function, the item $i$'s attribute embedding $\hat{\bm{t}}_i$ is able to indicate what attributes the item~$i$ has.

We then introduce how to construct the user $u$'s specific preference distribution~$\mathcal{G}(\bm{\mu}_u^s, \bm{\Sigma}_u^s)$ from the general one based on the general-specific transformation. Differently from the general preference distribution, each dimension in the specific preference distribution refers to a specific item's attribute. Therefore, the mean vector $\bm{\mu}^s_u \in \mathbb{R}^{|\mathcal{A}|}$, i.e., the center of the user's specific preference, indicates what attributes that the user prefers. 
The covariance matrix $\bm{\Sigma}_u^s \in \mathbb{R}^{|\mathcal{A}| \times |\mathcal{A}|}$ stands for the relationships of these preferences.
Technically, we project the parameters of the general preference distribution, i.e., $\bm{\mu}^g_u$ and $\bm{\Sigma}^g_u$, into their specific counterparts as follows,
\begin{equation}
\label{equation_distri_trans}
\left\{
\begin{array}{lr}
\hat{\bm{\mu}}_u = \bm{W}_t\bm{\mu}^g_u, \\
\hat{\bm{\Sigma}}_u = \bm{W}_t\bm{\Sigma}^g_u\bm{W}_t^\mathsf{T}.
\end{array}
\right.
\end{equation}	

In this manner, the transformed parameters, i.e., $\hat{\bm{\mu}}_u$ and $\hat{\bm{\Sigma}}_u$, are the mean vector and covariance matrix of the specific preference distribution, respectively, i.e., $\hat{\bm{\mu}}_u=\bm{\mu}^s_u, \hat{\bm{\Sigma}}_u=\bm{\Sigma}^s_u$.
The rationality proof is given as follows,
\begin{itemize}[leftmargin=6.5mm]
	\item \textit{\textbf{Argument 1.} $\hat{\bm{\mu}}_u$ is the mean vector of the user~$u$'s specific preference distribution.}
	
	\item \textit{\textbf{Proof 1.}} \textit{Referring to the mathematical definition of the mean vector, and the additivity and homogeneity of the linear mapping, we have,}
\begin{small}
	\begin{equation}\nonumber
		\begin{aligned}	
	\hat{\bm{\mu}}_u &= \bm{W}_t\bm{\mu}^g_u = \bm{W}_t (\frac{1}{|\mathcal{I}_u|}\sum_{i \in \mathcal{I}_u}\bm{f}_i) = \frac{1}{|\mathcal{I}_u|}\bm{W}_t\sum_{i \in \mathcal{I}_u}\bm{f}_i\\	
	&  = \frac{1}{|\mathcal{I}_u|}\sum_{i \in \mathcal{I}_u}\bm{W}_t\bm{f}_i = \frac{1}{|\mathcal{I}_u|}\sum_{i \in \mathcal{I}_u} 
	\hat{\bm{t}}_i:=\bm{\mu}^s_u. \qquad\Box 
	\end{aligned}
	\end{equation}
\end{small}

\item \textit{\textbf{Argument 2.} $\hat{\bm{\Sigma}}_u$ is the covariance matrix of the user~$u$'s specific preference distribution.}  
\item \textit{\textbf{Proof 2.}}
\textit{Referring to the mathematical definition of the covariance matrix, and the additivity and homogeneity of the linear mapping, we have,}
\begin{small}
	\begin{align}
	\hat{\bm{\Sigma}}_u&=\bm{W}_t\bm{\Sigma}^g_u\bm{W}_t^\mathsf{T} \notag =  \bm{W}_t (\frac{1}{|\mathcal{I}_u|}\sum_{i\in\mathcal{I}_u}(\bm{f}_i-\bm{\mu}^g_u)(\bm{f}_i-\bm{\mu}^g_u)^\mathsf{T}) \bm{W}_t^\mathsf{T}\notag\\
	&=\frac{1}{|\mathcal{I}_u|}\sum_{i \in \mathcal{I}_u} \bm{W}_t(\bm{f}_i-\bm{\mu}^g_u)(\bm{f}_i-\bm{\mu}^g_u)^\mathsf{T}\bm{W}_t^\mathsf{T}  = \frac{1}{|\mathcal{I}_u|}\sum_{i \in \mathcal{I}_u} (\bm{W}_t(\bm{f}_i-\bm{\mu}^g_u))(\bm{W}_t(\bm{f}_i-\bm{\mu}^g_u))^\mathsf{T} \notag\\ &= \frac{1}{|\mathcal{I}_u|}\sum_{i \in \mathcal{I}_u} (\bm{W}_t\bm{f}_i-\bm{W}_t\bm{\mu}^g_u)(\bm{W}_t\bm{f}_i-\bm{W}_t\bm{\mu}^g_u)^\mathsf{T} = \frac{1}{|\mathcal{I}_u|}\sum_{i \in \mathcal{I}_u} (\hat{\bm{t}}_i-\hat{\bm{\mu}}_u)(\hat{\bm{t}}_i-\hat{\bm{\mu}}_u)^\mathsf{T} \notag\\
	&= \frac{1}{|\mathcal{I}_u|}\sum_{i \in \mathcal{I}_u} (\hat{\bm{t}}_i-\bm{\mu}^s_u)(\hat{\bm{t}}_i-\bm{\mu}^s_u)^\mathsf{T} :=\bm{\Sigma}^s_u.\notag \qquad\Box 
	\end{align} 
\end{small} 
\end{itemize}

After constructing the user's specific preference distribution, we define the specific preference~$p_{ui}^s$ of the user~$u$ toward the item~$i$ as the probability density of the item $i$'s attribute embedding $\hat{\bm{t}}_i$ in the similar form as follows,
\begin{equation}
	\label{equation_p^s_ui}
		p^s_{ui} = P(\hat{\bm{t}}_i|\bm{\mu}_u^s, \bm{\Sigma}_u^s) \\
		= \frac{1}{\sqrt{2\pi|\bm{\Sigma}_u^s|}} \exp \big(-\frac{1}{2}(\hat{\bm{t}}_i-\bm{\mu}_u^s)^\mathsf{T}({\bm{\Sigma}_u^s})^{-1}(\hat{\bm{t}}_i-\bm{\mu}_u^s)\big).
\end{equation}

\begin{figure*}[!t]
	\centering
	\includegraphics[width=0.85\textwidth]{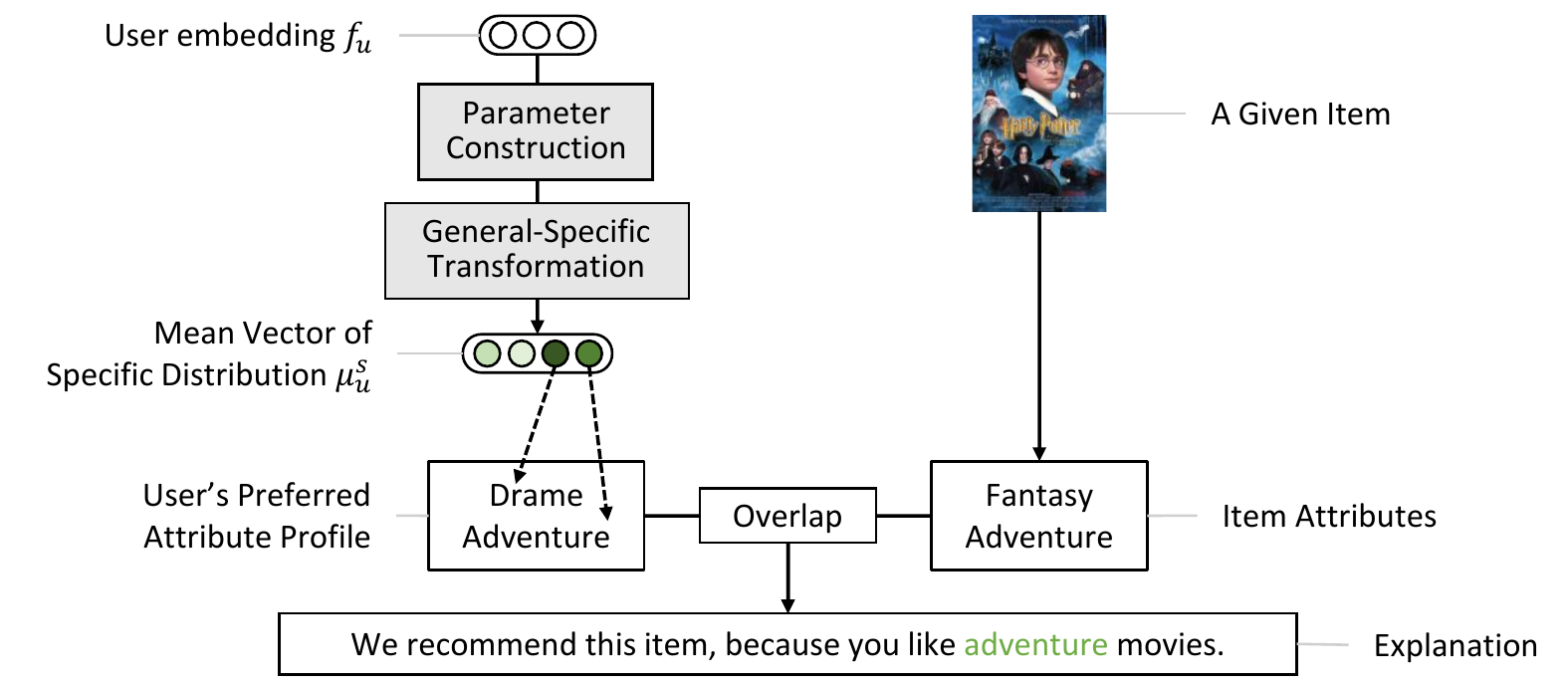}
	\caption{Illustration of the explanation production of the proposed DUPLE. We first derive the user's preferred attribute profile with the mean vector of the learned user's specific preference distribution. We then use the overlapped attributes between the learned attribute profile and the item's attributes as the explanation.}
	\label{figure_explanation_generation}
\end{figure*}

\subsection{Explanation Production}
\label{section_explanation}
In addition to the item recommendation, DUPLE can also explain why to recommend an item to a user as a byproduct, as shown in Fig.~\ref{figure_explanation_generation}. In particular, in the specific preference distribution, each dimension refers to a specific attribute and the mean vector indicates the center of the user's specific preference. Accordingly, we can infer what attributes the user likes and summarize a preferred attribute profile for the user. Technically, suppose that the user~$u$ prefers $r$ attributes, then we can define the preferred attribute profile $\mathcal{A}_u$ as follows,
\begin{equation}
\label{equation_profile}
\left\{
\begin{array}{lr}
e_1,e_2,...,e_r = \arg \max_r (\bm{\mu}^s_u), \bm{\mu}^s_u \in \mathbb{R}^{|\mathcal{A}|},\\
\mathcal{A}_u=\{a_{e_1},a_{e_2},...,a_{e_r}\},\\
\end{array}
\right.
\end{equation}
where $\arg \max_r (\cdot)$ is the function returning the indices of the $r$ largest elements in the vector. $a_{e_j}$ is the $e_j$-th attribute in $\mathcal{A}$.
After building the preferred attribute profile of the user~$u$, for a given recommended item~$i$ associated with its attributes~$\mathcal{A}_i$, we can derive the reason why the user likes the item by checking the overlap between $\mathcal{A}_u$ and $\mathcal{A}_i$. Suppose that $\mathcal{A}_u \cap \mathcal{A}_i=\{a_{e_j},a_{e_k}\}$. Then we can provide the explanation for user $u$ as ``you may like the item $i$ for its attributes $a_{e_j}$ and $a_{e_k}$''. 

Besides, it is worth mentioning that the diagonal elements of the covariance matrix~$\bm{\Sigma}_u^s$ in the specific preference distribution can capture the significance of user's different preferences to attributes. Specifically, if the value of the diagonal element in one dimension is small, its tiny change will sharply affect the prediction of the user-item interaction, and we can infer that the user's preference to the attribute in the corresponding dimension is strong.

\subsection{Optimization}
\label{section_interaction}
\begin{algorithm}[t]
	\caption{Dual Preference Distribution Learning Framework (DUPLE).}
	\label{algorithm_training}
	\begin{algorithmic}[1]
		\REQUIRE The sets of users $\mathcal{U}$ and items $\mathcal{I}$. The set of attributes~$\mathcal{A}_i$ of each~$i \in \mathcal{I}$. The training set~$\mathcal{D}$. The number of the optimization step $N$.
		\ENSURE The parameters~$\boldsymbol{\Theta}$ of DUPLE.\\
		\STATE Initialize the user embedding~$\bm{f}_u, u\in \mathcal{U}$, and item embedding~$\bm{f}_i, i\in \mathcal{I}$. \\
		Initialize the model parameters~$\boldsymbol{\Theta}$.
		\FOR {$n$ in $[1, \cdots, N]$}
		\STATE Randomly draw a mini-batch of training triplets~$(u,i,j)$ from~$\mathcal{D}$.
		\STATE Calculate the mean vector~$\bm{\mu}_u^g$ and covariance matrix~$\bm{\Sigma}_u^g$ of the general preference distribution through Eqn.~(\ref{equation_mu}) and (\ref{equation_sigma}), respectively.   \\
		\STATE Calculate the mean vector~$\bm{\mu}_u^s$ and covariance matrix~$\bm{\Sigma}_u^s$ of the specific preference distribution through Eqn.~(\ref{equation_distri_trans}). 
		 \\
		\STATE Predict the user-item interaction $p_{ui}$($p_{uj}$) through Eqn.~(\ref{equation_pui}).\\
		\STATE Update the parameters through Eqn.~(\ref{equation_L}): \\
		\centering $\boldsymbol{\Theta} \leftarrow \boldsymbol{\Theta}-\eta \frac{\partial \mathcal{L}}{\partial\boldsymbol{\Theta}} $
		\ENDFOR
	\end{algorithmic}
\end{algorithm}

The learned general and specific preference distributions are expected to assign a higher probability for the items that the user historically interacted item, and vice versa. Thus, regarding the optimization of the proposed DUPLE method, we build the following training set according to the Bayesian personalized ranking mechanism~\cite{RendleFGS09},
\begin{equation}
\label{eqn_trainin_set}
    \mathcal{D} = \{(u,i,j)|u\in\mathcal{U},i\in \mathcal{I}_u, j\in \mathcal{I}\setminus \mathcal{I}_u\},
\end{equation}
where the training triplet~$(u,i,j)$ indicates that the user~$u$ prefers item $i$ to the item~$j$. 

Ultimately, based on our constructed training set in Eqn.~(\ref{eqn_trainin_set}), we define the objective function for the DUPLE model as follows,
\begin{equation}
	\label{equation_L}
	\mathcal{L} = \min_{\boldsymbol{\Theta}} \big( \mathcal{L}_t + \sum_{(u,i,j) \in \mathcal{D}} -\log(\frac{p_{ui}}{p_{ui}+p_{uj}})\big),
\end{equation}
where $p_{ui}$ ($p_{uj}$) is the user-item interactions between the user~$u$ and item~$i$ ($j$) defined in Eqn.~(\ref{equation_pui}). $\mathcal{L}_t$ is the loss function of the general-specific transformation defined in Eqn.~(\ref{equation_Lt}). 
$\boldsymbol{\Theta}$ refers to the set of to-be-learned parameters of the proposed framework. The detailed training process of DUPLE is summarized in Algorithm~\ref{algorithm_training}.
	
\section{Experiments}
In this section, we first introduce the dataset details and experimental settings in Subsections~\ref{dataset_section} and \ref{experimental_settings}, respectively. 
And then, we conduct extensive experiments by answering the following research questions:
\begin{enumerate}
    \item Does DUPLE outperform the state-of-the-art methods?
    \item How do the different variants of the Gaussian distribution perform?
    \item What are the learned relations of the user's preferences?
    \item How is the explainable ability of DUPLE for the item recommendation?
\end{enumerate}

\subsection{Dataset and Pre-processing}
\label{dataset_section}
To verify the effectiveness of DUPLE, we adopted six public datasets with various sizes and densities: the Women's Clothing, Men's Clothing, Cell Phones \& Accessories, MovieLens-small, MovieLens-1M, and MovieLens-10M. The former three datasets are derived from Amazon Product dataset~\cite{McAuleyTSH15}, where each item is associated with a textual description. The latter three datasets are released by MovieLens dataset~\cite{HarperK16}, where each movie has the title, publication year, and genre information. User ratings of all the six datasets range from 1 to 5. To gain the reliable preferred items of each user, following the studies~\cite{LiuZLLGLJ18, 0020X0MZZT20}, we only kept the user's ratings that are larger than 3. 
Meanwhile, similar to the studies~\cite{KangM18, GeXLFSZ20}, for each dataset, we filtered out users and items that have less than~10 interactions to ensure the dataset quality.

\begin{table*}[t]\small
	\centering
	\caption{Statistics of the six public datasets (after preprocessing), including the numbers of users (\#user), items (\#item), their interactions (\#rating), and attributes (\#attribute), as well as the density of the dataset.}
	\label{table_dataset}  
	\setlength{\tabcolsep}{2.5mm}{
		\begin{tabular}{c|rrrrr}
			\hline
			&\multicolumn{5}{c}{Amazon Product Dataset}\\
			\cline{2-6}
			 &\#user&\#item & \#rating & \#attribute & density \\
			\hline	
			Women's Clothing &19,972&285,508&326,968&1,095&0.01\% \\	
			Men's Clothing &4,807&43,832&70,723&985&0.03\%\\
			Cell Phone \& Accessories&9,103&51,497&132,422&1,103&0.03\% \\
			\hline	
			&\multicolumn{5}{c}{MovieLens Dataset}\\
			\cline{2-6}
			 &\#user&\#item & \#rating & \#attribute & density \\
			\hline 
			MovieLens-small &579&6,296&48,395&698&1.33\%\\		
			MovieLens-1M &5,950&3,532&574,619&543&2.73\%\\
			MovieLens-10M &66,028&10,254&4,980,475&446&0.74\%\\	
			\hline
	\end{tabular}}
	\vspace{-0.0cm}	
\end{table*}

Since there is no attribute annotation in all the datasets above, following the studies~\cite{WangWJY18, WuQLWZL19}, we adopted the high-frequency words in the item's textual information as the ground truth attributes of items. Specifically, for each dataset, we regarded the words that appear in the textual description of more than 0.1\% of items in the dataset as high-frequency words. Notably, the stopwords (like ``the'') and noisy characters (like ``/'') are not considered. The final statistics of the datasets are listed in Table~\ref{table_dataset}, including the numbers of users (\#user), items (\#item), their interactions (\#rating), and attributes (\#attribute), as well as the density of the dataset. Similar to the study~\cite{WangJZ0XC20}, we calculated the dataset density by the formula~$\frac{\#
rating}{\#user \times \#item}$. 

\subsection{Experimental Settings}
\label{experimental_settings}

\textbf{Data Split.} We adopted the widely used leave-one-out evaluation~\cite{HeLZNHC17, GeXLFSZ20} to split the training, validation, and testing sets. In particular, for each user, we randomly selected an item from his/her historical interacted items for validation and testing, respectively, and left the rest for training. In the validation and testing, in order to avoid the heavy computation on all user-item pairs, following the studies~\cite{HeLZNHC17, GeXLFSZ20}, we composed the candidate item set by one \mbox{ground-truth} item and 100 randomly selected negative items that have not been interacted by the user. 

\begin{figure*}[!t]
	\centering
	\includegraphics[width=\textwidth]{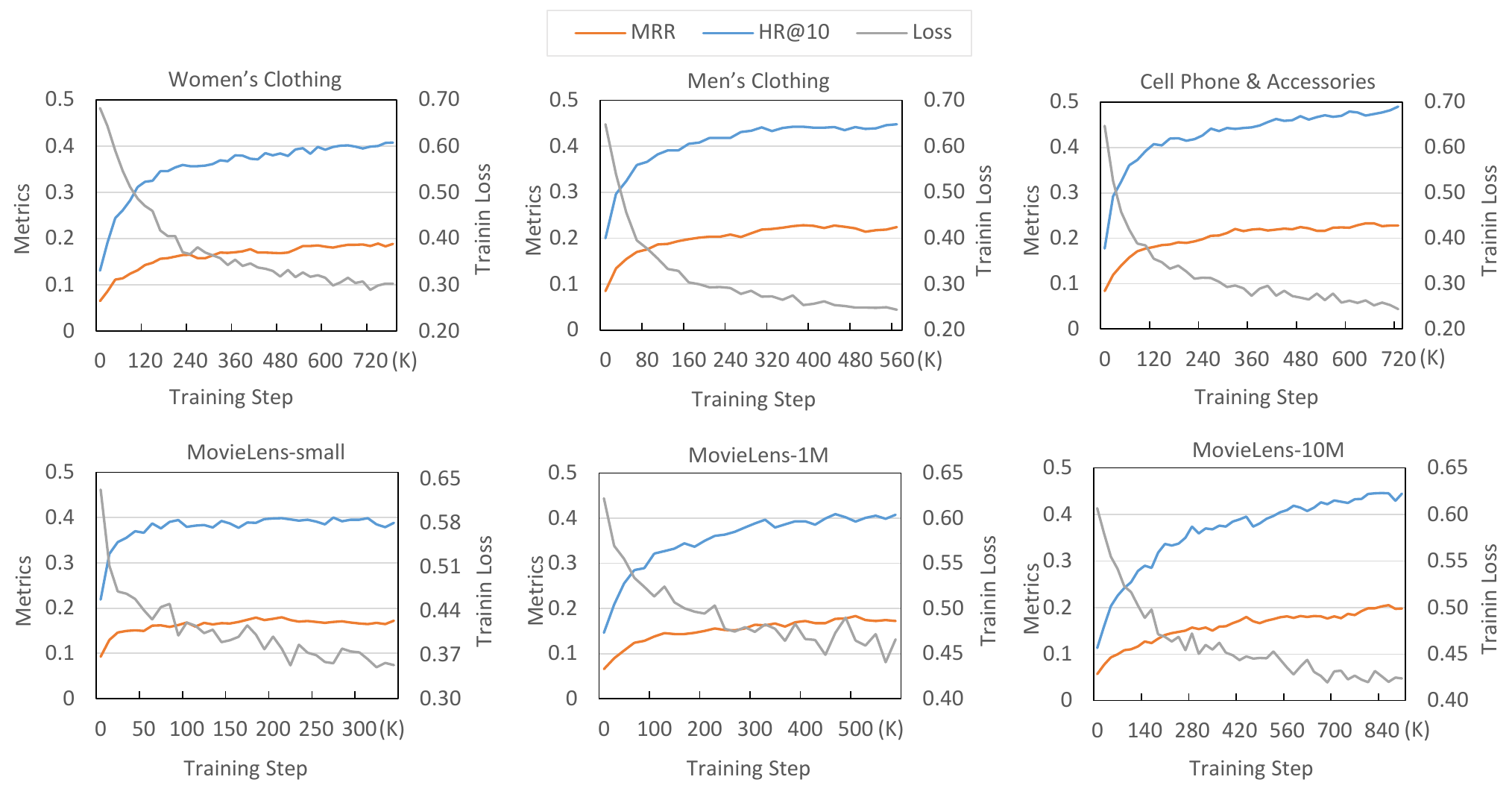}
	\caption{The training curves of our DUPLE model on the six datasets. The training loss is shown in grey line corresponding to the right longitudinal axis. MRR and HR@10 on the validation set are shown in orange and blue lines, respectively, corresponding to the left longitudinal axis.}
	\label{figure_curve}
\end{figure*}

\textbf{Evaluation Metrics.} 
We adopted the Area Under Curve (AUC), Mean Reciprocal Rank (MRR), Hit Rate (HR@10) and Normalized Discounted Cumulative Gain (NDCG@10) truncated the ranking list at 10 to comprehensively evaluate the performance. In particular, AUC indicates the classification ability of the model in terms of distinguishing the user's likes and dislikes. MRR, HR@10, and NDCG@10 reflect the ranking ability of the model in terms of the top-N recommendation.

\textbf{Implementation Details.}
Following the study~\cite{GeXLFSZ20},  we unified the dimension $D$ of the item and user embedding as 64. 
Besides, to gain the powerful embedding of the user and item, we added a two-layer perceptron to transform the raw embeddings before feeding them into the network.
We used the random normal initialization for the parameters and trained them by Adam optimizer~\cite{KingmaB14} with a learning rate of $10^{-4}$ and batch size of $256$. To derive the training set~$\mathcal{D}$ introduced in Eqn.~(\ref{eqn_trainin_set}), following studies~\cite{SongFLLNM17, Wang00LC19}, for each pair of user~$u$ and an interacted item~$i$ in the training batch, we randomly sampled an item that has not been interacted by the user~$u$ as the item~$j$.
For different dataset, the number of the optimization steps $N$ is different. This is because a large dataset needs more steps to converge. We showed the curves of the training loss in Eqn.~(\ref{equation_L}) and metrics (i.e., MRR and NDCG@10) on the validation set on the six datasets in Fig.~\ref{figure_curve}.
For our method, we tuned the trade-off parameter~$\lambda$ in Eqn.(\ref{equation_pui}) from $0$ to $1$ with the stride of $0.1$ for each dataset, and the dimension $D'$ in Eqn.(\ref{equation_sigma}) from $[2,4,8,16,32]$.

\subsection{Comparison of Baselines (RQ1)}
\label{performance_compare}
We compared the proposed DUPLE method with the following baselines.
\begin{itemize}[leftmargin=8mm]
	\item \textbf{BPR~\cite{RendleFGS09}.} Bayesian personalized ranking (BPR) is one of the most widely used methods for the top-$N$ recommendation. 
	It represents the users and items with feature vectors and introduces a personalized ranking criterion for the optimization, which aims to yield a larger similarity for a user and a positive item as compared to a user and a negative item.
	\item \textbf{AMR~\cite{TangDHYTC20}.} To enhance the recommending robustness, this approach involves the adversarial learning onto the BPR model. Specifically, it trains the network to defend an adversary by adding perturbations to the item image.
	\item \textbf{DVMF~\cite{ShenYLZZLX21}}. This model is a distribution-based method for the click prediction task, which represents users and items with Gaussian distributions, respectively. DVMF designs a densely-connect multi-Layer perceptron (D-MLP) to produce the parameters of the Gaussian distribution based on the randomly initialized embedding, and utilizes the variational inference to measure the user-item rating. We replaced its objective function of the classification with the ranking loss as the same as our work. It is worth noting that this modification boosts the performance of DVNE in the context of top-N recommendation. 
	\item \textbf{ex-DVMF}. This method is an extension of DVMF. To explore the user's specific preference to item attributes, it further involves the item attribute embedding to enrich the item embedding. Specifically, the bag-of-words attribute embedding of the item is first mapped into the same dimension of the item embedding, and then the summation of the two embeddings is adopted as the enriched item embedding.
	\item \textbf{AMCF~\cite{PanLLZ20}.} The attentive multitask collaborative filtering (AMCF) method adds the item's attributes into the matrix factorization by learning a projection between the item embedding and attribute embedding. It is an extension of the BPR method, which utilizes the weighted-summation of item attributes as a supervision to learn the item embedding. According to the weights, AMCF can determine the user's specific preferences to different attributes. It can be regarded as adding a multi-task learning into the BPR method, where one task is to rank the items and another is to predict the item attribute. 
	\item \textbf{NARRE~\cite{ChenZLM18}.} This method is designed for the click prediction, which adds the item's attributes into the matrix factorization. It enriches the item embedding with the weighted-summation of all the word embedding of the item's attributes. 
	Ultimately, based on the learned weights for the attribute embeddings, NARRE can capture the user's specific preferences to attributes. We also replaced its objective function with the ranking loss to boost its performance in the top-N recommendation.
\end{itemize}

\begin{table*}[t]\small
	\centering
	\renewcommand{\arraystretch}{1.2}
	\caption{Comparison results on the six datasets. The best and second-best results are in bold and underlined, respectively. \%impro is the relative improvement of the proposed DUPLE compared to the strongest baseline.} 
	\label{table_comparison}
	\setlength{\tabcolsep}{1.2mm}{
		\begin{tabular}{c|c|cccc|c|c|cccc}	
			\hline\hline
			&\multirow{2}*{Method}&\multicolumn{4}{c|}{Metrics}&&\multirow{2}*{Method}&\multicolumn{4}{c}{Metrics}\\ 
			\cline{3-6}\cline{9-12}
			&&AUC & MRR & HR@10 & NDCG@10 &&& AUC & MRR & HR@10 & NDCG@10 \\
			\hline  
			\multirow{8}*{\rotatebox{90}{Women's Clothing}}&BPR &54.01&6.73&13.43&6.49&\multirow{8}*{\rotatebox{90}{MovieLens-Small}}&BPR&72.04&12.12&29.36&13.49\\
			&AMR &55.94&7.79&16.73&11.25&&AMR &74.66&12.66&32.70&13.85\\
		    &DVMF &58.45&8.32&19.80&12.41&&DVMF &73.48&13.07&33.16&15.98\\
			\cline{2-6}\cline{8-12} 
		    &ex-DVMF &70.09&11.72&27.33&13.33&&ex-DVMF &\underline{77.54}&\underline{15.30}&\underline{37.01}&\underline{18.28}\\ 
			&AMCF &62.18&9.85&22.14&10.86&&AMCF &76.06&11.91&35.15&17.04\\
			&NARRE&\underline{74.93}&\underline{17.95}&\underline{39.12}&\underline{21.13}&&NARRE&76.30&13.23&36.84&17.41\\
			&DUPLE&\textbf{77.48}&\textbf{20.35}&\textbf{42.43}&\textbf{23.60}&&DUPLE&\textbf{79.07}&\textbf{16.43}&\textbf{38.86}&\textbf{20.80}\\
			\cline{2-6}\cline{8-12}
			&\%impro&+3.40&+13.37&+8.46&+11.68&&\%impro&+1.97&+7.38&+4.99&+13.78\\
			\hline \hline

			\multirow{8}*{\rotatebox{90}{Men's Clothing}}&BPR &54.40&6.50&12.81&6.12&\multirow{8}*{\rotatebox{90}{MovieLens-1M}}&BPR &73.95&15.26&33.83&17.23\\
			&AMR &55.99&7.35&14.54&7.09&&AMR &74.73&15.37&34.21&18.52\\
			&DVMF &59.11&8.97&20.65&13.27&&DVMF &77.91&17.15&39.00&20.34\\
			\cline{2-6}\cline{8-12}  
			&ex-DVMF &66.51&11.24&25.02&12.54&&ex-DVMF &\underline{79.20}&\underline{18.59}&\underline{40.97}&\underline{21.91}\\ 
			&AMCF &60.10&9.46&20.44&10.20&&AMCF &77.90&15.42&38.18&18.79\\
			&NARRE&\underline{72.49}&\underline{18.30}&\underline{38.00}&\underline{21.20}&&NARRE&79.06&17.45&40.23&20.55\\  
			&DUPLE&\textbf{76.00}&\textbf{19.35}&\textbf{40.95}&\textbf{23.62}&&DUPLE&\textbf{79.52}&\textbf{18.80}&\textbf{42.85}&\textbf{22.28}\\
			\cline{2-6}\cline{8-12}
			&\%impro&+4.84&+5.73&+7.76&+11.41&&\%impro&+0.40&+1.12&+4.58&+1.68\\
			\hline \hline

			\multirow{8}*{\rotatebox{90}{Cell Phone \& Accessories}}&BPR  &66.97&13.06&28.75&14.89&\multirow{8}*{\rotatebox{90}{MovieLens-10M}}&BPR  &76.19&15.17&34.97&17.47\\ 
			&AMR &67.12&13.60&30.02&15.65&&AMR &78.12&16.15&38.47&18.39\\
			&DVMF &68.71&15.55&30.10&16.28&&DVMF&84.04& 22.31&49.07&26.72 \\
			\cline{2-6}\cline{8-12} 
			&ex-DVMF  &75.88&15.29&36.59&18.31&&ex-DVMF&\textbf{84.74}&\underline{23.34}&\underline{50.97}&\underline{28.03}  \\ 
			&AMCF &69.61&13.45&30.85&15.61&&AMCF &80.50&16.68&40.91&20.68\\
			&NARRE&\underline{77.53}&\underline{18.70}&\underline{41.15}&\underline{22.11}&&NARRE&83.90&22.22&49.29&26.41\\  
			&DUPLE&\textbf{80.65}&\textbf{20.38}&\textbf{45.47}&\textbf{25.14}&&DUPLE & \underline{84.70}&\textbf{23.74}&\textbf{51.67}&\textbf{28.10}\\
			\cline{2-6}\cline{8-12}
			&\%impro&+4.02&+8.98&+10.49&+13.70&&\%impro&--0.04&+1.71&+1.37&+0.07\\
			\hline\hline

	\end{tabular}}
\end{table*}

It is worth noting that the first three baselines, i.e., BPR, AMR, and DVMF, only focus on the user's general preference, termed as single-preference-based (\mbox{SP-based}) method. The rest methods, i.e., ex-DVMF, NARRE, AMCF, and our proposed DUPLE consider both the general and specific preferences during recommending items, termed as dual-preference-based (\mbox{DP-based}) method. 
For each method, we reported its average results of three runs with different random initialization of model parameters. 
Table~\ref{table_comparison} shows the performance of baselines and our proposed DUPLE method on the six datasets. The best and second-best results are in bold and underlined, respectively. The row ``\%improv''  indicates the relative improvement of DUPLE over the best results of baselines. 
From Table~\ref{table_comparison}, we have the following observations: 

\begin{enumerate}[leftmargin=8mm]
\item  DUPLE outperforms all the baselines in terms of almost metrics across different datasets, which demonstrates the superiority of our proposed framework over existing methods. 
This may be due to the fact that by learning the user's dual preferences (i.e., general and specific preferences) with the probabilistic distributions, DUPLE is capable of exploring the relationships of the user's difference preferences, whereby gains the better performance. 
\item The DP-based methods (i.e., ex-DVMF, NARRE, AMCF, and DUPLE) gain the better performance than the SP-based methods (i.e., BPR, AMR, and DVMF) on average. It proves that jointly learning the user's general and specific preferences helps to better understand the user overall preferences and gains a better recommendation performance. Besides, we found that in datasets from Amazon, the improvements of the DP-based methods over SP-based methods are larger than improvements in datasets from MovieLens. This may be because that datasets from Amazon are sparse so that the user and item embeddings cannot be learned well with the limited user-item interacted data. In these cases, engaging the information of the item attribute helps more to understand the item properties and better learn the user's preferences.
\item Among SP-based methods, AMR outperforms BPR on all the datasets. This indicates that adding perturbations makes the network more robust and has a better generalization ability. Besides, DVMF that utilizes the distribution to represent user's preferences outperforms the other SP-based methods, i.e., BPR and AMR. This proves that a distribution has the better descriptive power to represent the user's preferences than a vector. 
\item. On the one hand, AMCF outperforms BPR with a large margin in all datasets, demonstrating the benefit of modeling the user's specific preference to attributes. On the other hand, AMCF performs worst among all DP-based methods. This may be because that this method only utilizes the item attributes as the supervision, ignoring the explicit modeling of the user's specific preferences. This suggests that it is better to directly model the general and specific preferences rather than only adding the multi-task learning.
\end{enumerate}

\subsection{Comparison of Variants (RQ2)}
\label{distribution_analysis}
\begin{table*}[t]\small
	\centering
	\renewcommand{\arraystretch}{1.1}
	\caption{The results of the comparison of different variants of DUPLE. DUPLE-iden and DUPLE-diag set the covariance matrix of the preference distributions are identity and diagonal matrices, respectively. The best results are highlighted in bold.} 
	\label{table_ablation}
	\setlength{\tabcolsep}{1.3mm}{
		\begin{tabular}{c|c|p{1.2cm}<{\centering}p{1.2cm}<{\centering}p{1.2cm}<{\centering}p{1.4cm}<{\centering}}	
			\hline\hline
			&\multirow{2}*{Method}&\multicolumn{4}{c}{Metrics}\\
			\cline{3-6}
			& & AUC & MRR & HR@10 & NDCG@10 \\
			\hline
			
			\multirow{3}*{Women's Clothing}
			&DUPLE-iden &72.23&18.48&37.14&21.09\\
			&DUPLE-diag &75.19&18.43&40.14&21.22\\
			&DUPLE &\textbf{77.48}&\textbf{20.35}&\textbf{42.43}&\textbf{23.60}\\	
			\hline
			\multirow{3}*{Men's Clothing}			
			&DUPLE-iden &72.80&19.27&38.86&20.38\\
			&DUPLE-diag &75.76&18.16&\textbf{41.26}&22.16\\
			&DUPLE &\textbf{76.00}&\textbf{19.35}&40.95&\textbf{23.62}\\
			\hline
			\multirow{3}*{Cell Phone \& Accessories}
			&DUPLE-iden &77.84&19.49&41.10&24.00\\
			&DUPLE-diag &80.64&\textbf{20.99}&\textbf{46.64}&\textbf{25.65}\\
			&DUPLE &\textbf{80.65}&20.38&45.47&25.14\\
			\hline
			\multirow{3}*{MovieLens-Small}
			&DUPLE-iden &77.26&14.89&36.26&19.32\\
			&DUPLE-diag & 78.33&15.13&36.09&18.11\\
			&DUPLE 	&\textbf{79.07}&\textbf{16.43}&\textbf{38.86}&\textbf{20.80}\\
			\hline
			\multirow{3}*{MovieLens-1M}
			&DUPLE-iden &76.38&16.56&37.65&19.57\\
			&DUPLE-diag &76.53&15.66&36.87&17.57\\
			&DUPLE &\textbf{79.52}&\textbf{18.80}&\textbf{42.85}&\textbf{22.28}\\
			\hline
			\multirow{3}*{MovieLens-10M}
			&DUPLE-iden &79.16&17.57&40.32&20.93\\
			&DUPLE-diag &80.07&18.48&41.84&21.74\\
			&DUPLE &\textbf{84.70}&\textbf{23.74}&\textbf{51.67}&\textbf{28.10}\\
			\hline\hline
	\end{tabular}}
\end{table*}

In order to verify that the user's different preferences are related, we introduced the variant of our model, termed as \textbf{\mbox{DUPLE-diag}}, whose covariance matrices of the two distributions (i.e., $\bm{\Sigma}^g_u$ and $\bm{\Sigma}^s_u$) are set to be diagonal matrices, i.e., the off-diagonal elements of the covariance matrix are all zeros.
Besides, to further demonstrate that the user's different preferences contribute differently to predict the user-item interaction, we designed \textbf{DUPLE-iden} method that $\bm{\Sigma}^g_u = \bm{\Sigma}^s_u = {\rm \textbf{E}}$, where $\rm \textbf{E}$ is an identity matrix. Formally, according to Eqn.(\ref{equation_p^g_ui}) and Eqn.(\ref{equation_p^s_ui}), by omitting constant terms, the user $u$'s general and specific preferences can be simplified as $p^g_{ui} \propto e^{-\frac{1}{2}||\bm{f}_i-\bm{\mu}^g_u||_2}$ and $p^s_{ui} \propto e^{-\frac{1}{2}||\bm{\hat{t}}_i-\bm{\mu}^s_u||_2}$, respectively. Intuitively, the variant DUPLE-diag leverages Euclidean distance to measure the user preference, whose philosophy is as similar to the baselines we used. For each method, we reported the average result of three runs with different random initialization of the model parameters. The results are shown in Table~\ref{table_ablation} and the detailed analysis is given as follows:

\begin{enumerate}[leftmargin=8mm]
\item DUPLE outperforms the two variant methods with respect to almost all datasets, which demonstrates that it is necessary to learn the relationships and different contributions among the user's preferences to predict the user rating.
\item DUPLE-iden, whose covariance matrix is set to the identity matrix, performs worst in this comparison. Besides, the results of DUPLE-iden are comparable with the existing baselines on average. The reason behind this may be that with the identity covariance matrix setting, DUPLE-iden essentially represents the user's preferences by only a mean vector, which is the same as existing approaches that use vectorized embeddings, and thus achieves the similar performance. 
This proves that it is better to capture the user's preferences with probabilistic distribution compared to the vectorized embedding.

\item DUPLE-diag performs worse than DUPLE. The reason behind this may be that equipped with the diagonal covariance matrix, DUPLE-diag cannot capture the relationships among the user's different preferences. Differently, DUPLE can model such relationships by the off-diagonal elements of the covariance matrix, and thus better understand the user preferences.

\item In Cell Phone \& Accessories dataset, it is unexpected that DUPLE performs worse than DUPLE-diag on average. This may be attributed to that the user's different preferences to items in this category rarely interact with each other. For example, whether a user prefers \textit{black} phone will not be influenced by whether he/she prefers its \textit{LED-screen}. Therefore, leveraging extra parameters to capture these relations of user's preferences only decreases the performance. 
\end{enumerate}

\begin{figure*}[!t]
	\centering
	\includegraphics[width=\textwidth]{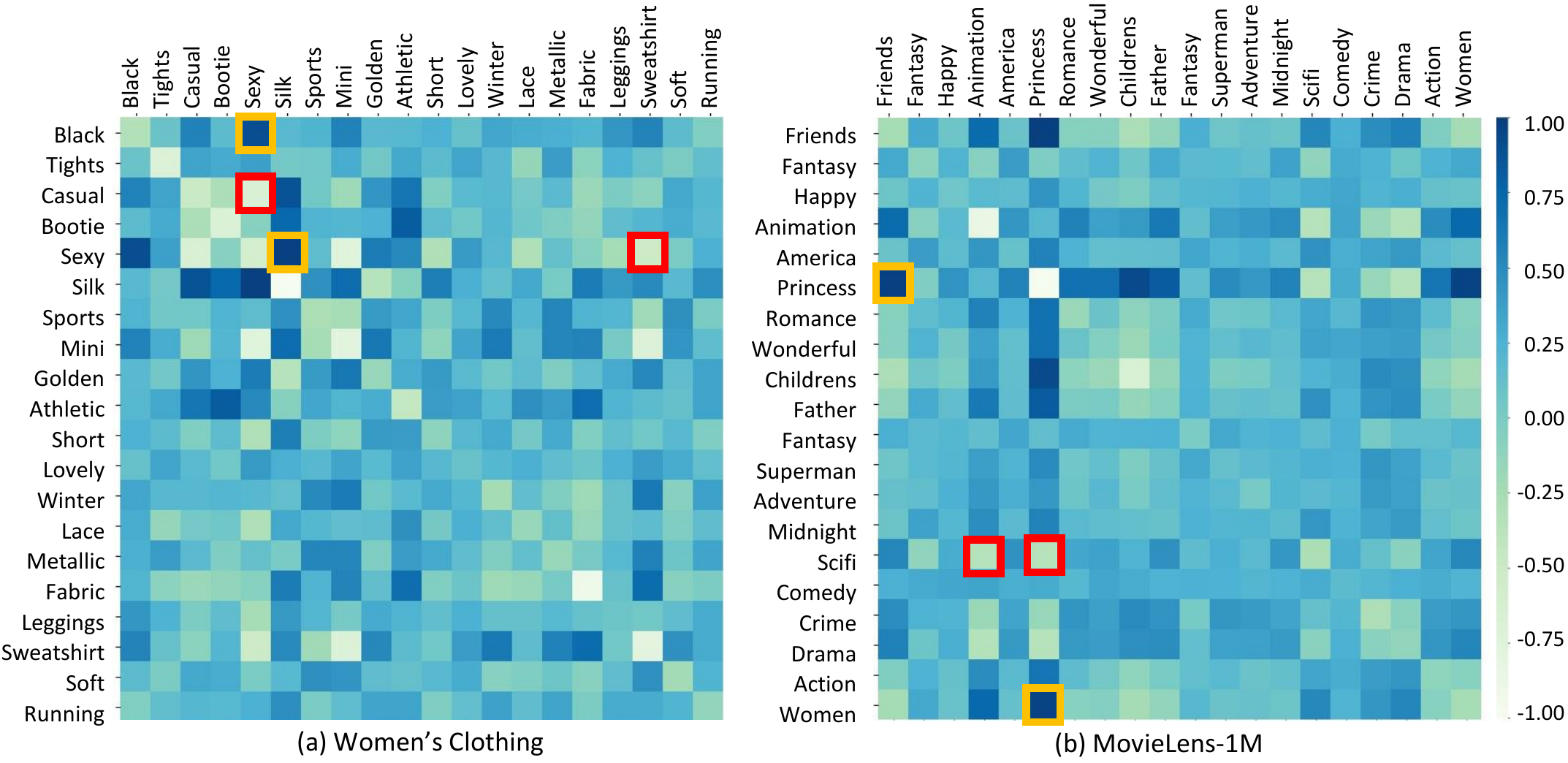}
	\caption{Visualization of the relationships among the user's preferences of a random user in Women's Clothing (a) and MovieLens-1M (b) datasets, respectively. The darker color indicates the higher relation, where the two highest and lowest preference pairs are surrounded by yellow and red boxes, respectively.}
	\label{figure_attribute_relationships}
\end{figure*}

\subsection{Visualization of User's Preferences Relationships (RQ3)}
\label{attribute}
In order to further gain a deeper insight of the relationships of the user's preferences, we visualized the learned covariance matrix of the specific preference distribution of a randomly selected user in the Women's Clothing and MovieLens-1M datasets, respectively. For clarity, instead of visualizing the relationships of the user's preferences to all the attributes, we randomly picked up 20 attributes and visualized their corresponding correlation coefficients in the covariance matrix of the specific preference distribution in each dataset by the heat map in Fig.~\ref{figure_attribute_relationships}. The darker blue color indicates that the user's preferences to the two attributes are higher related. We circled the four most prominent preference pairs, where two pairs with the highest relationship are surrounded by the yellow boxes, and two pairs with the lowest relationship by the red boxes.

From Fig.~\ref{figure_attribute_relationships}, we can see that in the dataset Women's Clothing, this user's preferences to attributes \textit{black} and \textit{sexy}, as well as attributes \textit{silk} and \textit{sexy}, are highly relevant, while those to attributes \textit{casual} and \textit{sexy}, as well as attributes \textit{sweatshirt} and \textit{sexy} are less relevant. This is reasonable as one user that prefers the sexy garments are more probably to like garments in black color, but hardly like casual garments. As for the MovieLens-1M dataset, the user's preference to attribute \textit{Princess} has the high relevance with the preferences to attributes \textit{Friends} and \textit{Women}, while the user's preferences to the attribute \textit{Scifi} are mutually exclusive with those to attributes \textit{Animation} and \textit{Princess}. These relationships uncovered by DUPLE also make sense. Overall, these observations demonstrate that the covariance matrix of one's specific preference distribution is able to capture the relationships of his/her preferences.

\begin{figure*}[!t]
	\centering
	\includegraphics[width=\textwidth]{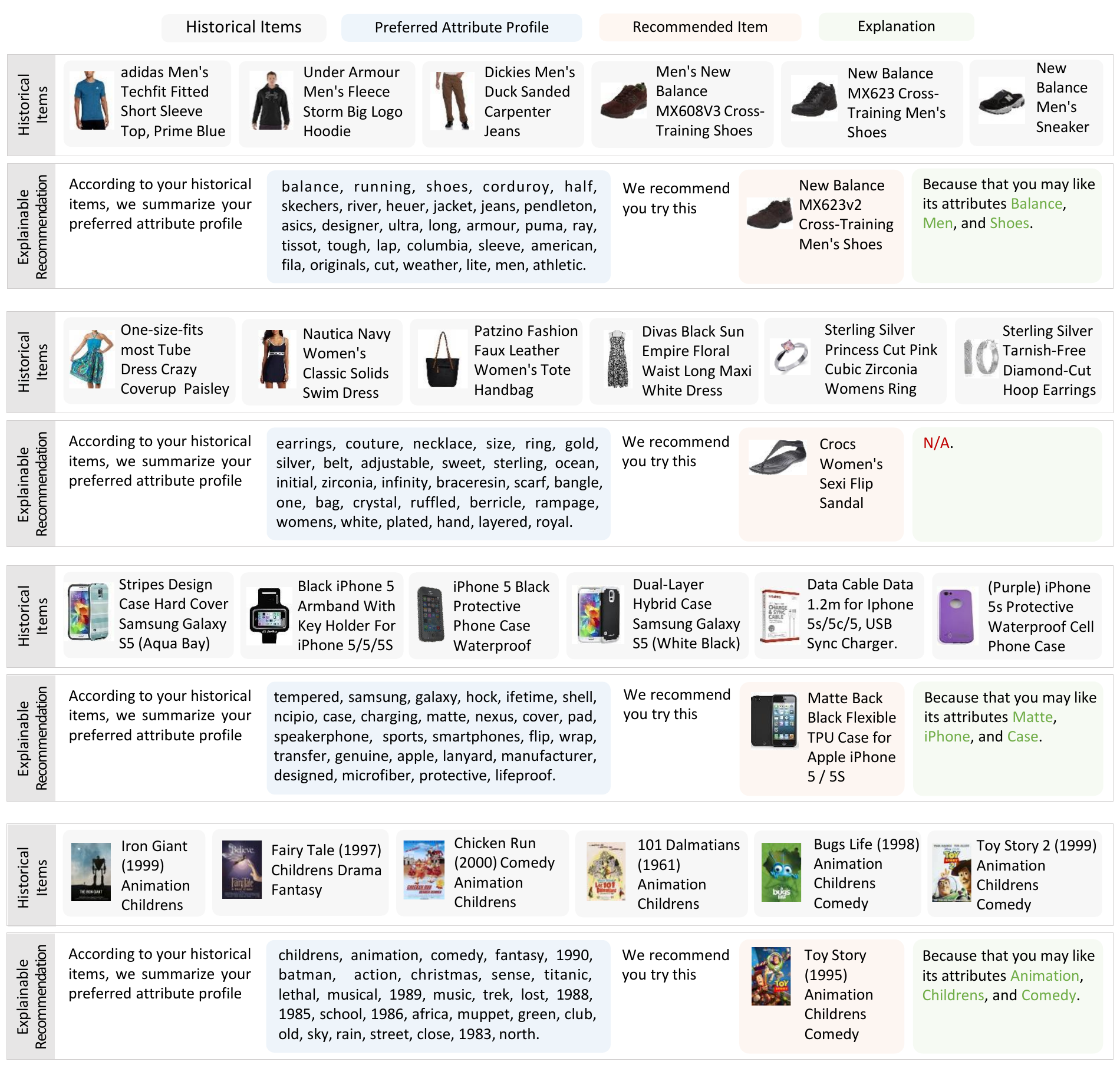}
	\caption{Examples of the explainable recommendation of the proposed DUPLE method. 
		Each example displays a user's historical preferred items, the preferred attribute profile summarized by DUPLE (the frontier position of the attribute means a bigger preference degree), and a recommended item with its exxplanations.}
	\label{figure_explanation}
\end{figure*}

\subsection{Explainable Recommendation (RQ4)}
\label{explainability}
To evaluate the explainable ability of our proposed DUPLE, we first provided examples of our explainable recommendation in Subsection~\ref{section_explainable_example}. We then conducted a subjective psycho-visual test to judge the explainability of the proposed DUPLE in Subsection~\ref{section_survey}. 

\subsubsection{Explainable Recommendation Examples}
\label{section_explainable_example}
Fig.~\ref{figure_explanation} shows four examples of our explainable recommendation. For each example, we list six randomly selected historical preferred items of the user for reference. We provided both images and textual descriptions of items to facilitate readers to learn the user's preferences. Meanwhile, we also provided the summarized user's preferred attribute profile, which is indispensable for producing the recommendation explanation. To be more specific, for each user, we calculated the user's specific preference distribution by our proposed DUPLE. We then derived the user's preferred attribute profile according to Eqn.~(\ref{equation_profile}). The users from the top to bottom in Fig.~\ref{figure_explanation} are from the datasets Men's Clothing, Women's Clothing, Cell Phone \& Accessories, and MovieLens-1M, respectively. 
From Fig.~\ref{figure_explanation}, we have the following observations.
\begin{enumerate}[leftmargin=8mm]	
	\item The summarized users' preferred attribute profiles in the center column are in line with the users' historical preferences. 
	For example, as for the first user, he has bought many sporty shoes and upper clothes, based on which we can infer that he likes sports and prefers the sporty style. These inferences are consistent with the preferred attributes DUPLE summarizes, e.g., DUPLE summarizes many brand of sports (\textit{skechers}, \textit{asics}, and \textit{columbia}).
	In addition, as for the 4-th user, he/she has watched several animations like \textit{Iran Giant} and \textit{Fairy Tales}. DUPLE correctly captures the user's preferences and summarizes his/her preferred attributes, including \textit{children's}, \textit{animation}, and \textit{comedy}.
	
	\item DUPLE is able to recommend the correct item and attach the reasonable explanations for the user. For example, the first user in Fig.~\ref{figure_explanation} has bought many New Balance (i.e., a sports brand) shoes historically. DUPLE has recommended the similar shoes of this brand and attached the explanation of ``The user like its attribute(s) \textit{Balance}, \textit{Men}, and \textit{Shoe}''. 
	
	\item Apparently, when there is no overlap between the attributes of the recommended item and the user's preferred attribute profile, DUPLE cannot provide the explanation. For example, for the second user in Fig.~\ref{figure_explanation}, DUPLE recommends the ``Crocs Women's Sexi Flip Sandal'' with no explanation. By checking the user's historically preferred items, we found this recommended sandal is highly compatible with the dresses in the user's historical interacted items. Thus, for such cases, although our model cannot provide the exact explanation, it still can correctly recommend the item according to the general preference. 
\end{enumerate}

\begin{figure}[!t]
	\centering
	\includegraphics[width=0.85\linewidth]{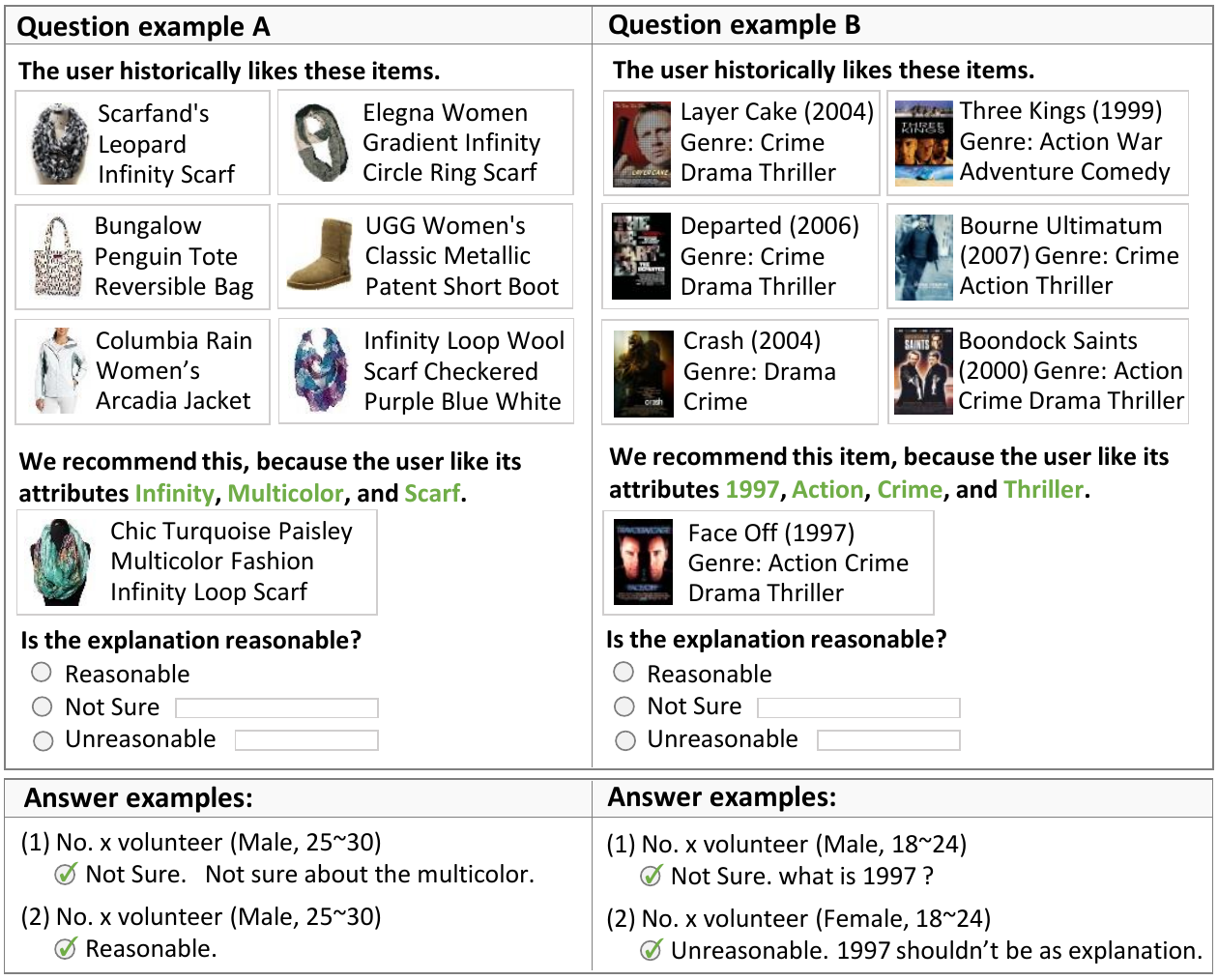}
	\caption{Two example questions in the subjective psycho-visual survey, each of which consists of the user's historical interacted items, a recommended item and explanation provided by DUPLE, and a judgment of the explanations. Two answers collected from volunteers are below the corresponding question, respectively.}
	\label{figure_survey_example}
\end{figure}

\subsubsection{Subjective Psycho-visual Test} 
\label{section_survey}
For the subjective psycho-visual test on judging the quality of the provided explanation of our DUPLE, we first designed a survey consisting of 12 questions (two questions from six datasets, respectively). Fig.~\ref{figure_survey_example} shows two question examples. As can be seen, each question consists of three parts: 6 items that a user historically likes, a newly recommended item with explanation, and a judgment of the explanation. For each question, volunteers first learned the user's preferences from the user's historical items and then made their judgment on the rationality of the explanation (choose from "Reasonable", "Not Sure", or "Unreasonable"). Meanwhile, if volunteers chose "Not Sure", or "Unreasonable", we required them to write down their decision reasons in the blank behind the option.

\begin{figure}[!t]
	\centering
	\includegraphics[width=0.8\linewidth]{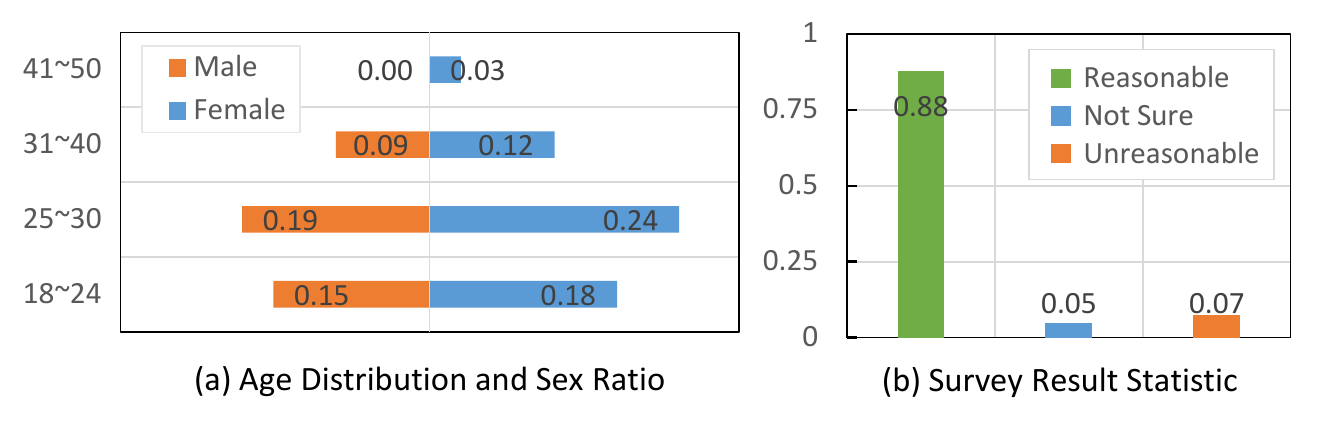}
	\caption{The information statistics of volunteers in the subjective psycho-visual survey (a). Survey results (b) of whether the volunteers think about the explanations provided by DUPLE.}
	\label{figure_survey_result}
\end{figure}

In total, we invited 126 volunteers to finish the above subjective psycho-visual test. The statistical information of the invited volunteers is listed in Fig.~\ref{figure_survey_result}~(a). The collected results of the psycho-visual test, shown in Fig.~\ref{figure_survey_result}~(b), is representative for the public, as male and female volunteers distributed homogeneously and their age ranged widely. Besides, we provided two examples of the volunteer's answers in Fig.~\ref{figure_survey_example} below the questions. Combining analyzing the survey results and answer examples, we have the following observations:

\begin{enumerate}[leftmargin=8mm]
	\item  Most volunteers (88\%) thought the explanations produced by DUPLE are reasonable. This demonstrates that our proposed DUPLE method can correctly provide the explanation for recommending an item to a user. 
	\item A few volunteers (5\%) are not sure about the explanations. This may be because that sometimes volunteers cannot derive the explicit cues in the user's historical preferred items for the certain attribute in the explanations. For example, as for the question A in Fig.~\ref{figure_survey_example}, some volunteers are not sure about explaining the recommended scarf with \textit{multicolor}. This may be because \textit{multicolor} has not explicitly appeared in the attributes of the user's historical preferred items, while the user's most historical preferred items are multi-color.
	\item 7\% volunteers thought the explanations are unreasonable. A part of volunteers thought it is unreasonable to explain the reason for recommending a movie with its publication year. For example, as for question B in Fig.~\ref{figure_survey_example}, DUPLE recommends the movie ``Face Off'' and explains that the user likes its attributes \textit{1997}. 
    Besides, some volunteers thought certain attributes should be treated as a whole as the recommended reason. For example, as for the first example in Fig.~\ref{figure_explanation}, DUPLE explains recommending the shoes with its attribute \textit{Balance}, while New Balance (a sports brand) should be treated as a whole. 
\end{enumerate}

\section{Conclusion and Future Work}
We propose a dual preference distribution learning framework (DUPLE), which captures the user's preferences from both the general item and specific attribute perspectives for a better recommendation. Different from existing approaches that represent the user and item as vectorized representations, DUPLE attempts to represent the user's preferences with the Gaussian distribution and then predict the user-item interaction by calculating the probability density at the item in the user's preference distribution. In this manner, DUPLE is able to explicitly model the relationships of the user's different preferences by the covariance matrix of the Gaussian distribution. Besides, the proposed DUPLE method can summarize a preferred attribute profile, depicting the item attributes that the user likes, based on which we can provide the explanation for a recommendation.
Quantitative and qualitative experiments conducted on six real-world datasets and the promising empirical results demonstrate the effectiveness and explainability of the proposed \mbox{DUPLE}.

Limitations of DUPLE include the two followings. 1) Currently, our method regards the high-frequency words in text descriptions as the item attributes, which may involve noises and hurt the specific preference learning. Thus, we plan to devise an attribute predictor that can automatically produce the attributes of an item from its text descriptions. 2) It is a little cumbersome that DUPLE needs to learn separate preference distributions to capture the user's general and specific preferences, and we need to tune the trade-off parameter to combine the user's general and specific preferences for different datasets. In the future, we plan to devise a flexible preference distribution that can jointly capture all kinds of the user's preferences to simplify the model.

\begin{acks}
This work is supported by the Shandong Provincial Natural Science Foundation, No.:ZR2022YQ59; and Alibaba Group through Alibaba Innovative Research Program.
\end{acks}

\bibliographystyle{ACM-Reference-Format}
\bibliography{ref.bib}


\end{document}